\documentstyle[epsfig,amssymb,twocolumn]{mn2e}

\newif\ifAMStwofonts
\title[Gravitational drag in MOND]
{Gaseous drag on a gravitational perturber in Modified Newtonian Dynamics
and the structure of the wake}
\author[S\'anchez-Salcedo ]
{F.~J.~S\'anchez-Salcedo\thanks{E-mail:jsanchez@astroscu.unam.mx}
\\
Instituto de Astronom\'{\i}a, Universidad Nacional Aut\'onoma
de M\'exico, Ciudad Universitaria, Apt.~Postal 70
264, \\
C.P. 04510, Mexico City, Mexico}

\begin{document}

\date{Accepted xxxx Month xx. Received xxxx Month xx; in original form
2008 August 6}
\pagerange{\pageref{firstpage}--\pageref{lastpage}} \pubyear{2007}
\maketitle

\label{firstpage}
\begin{abstract}
We calculate the structure of a wake generated by, and
the dynamical friction force on, a gravitational
perturber travelling through a gaseous medium of uniform density
and constant background acceleration $\bmath{g}_{\rm ext}$, 
in the context of Modified Newtonian
Dynamics (MOND). The wake is described as a linear superposition of two
terms. The dominant part displays the same structure
as the wake generated in Newtonian gravity scaled up by 
a factor $\mu^{-1}(g_{\rm ext}/a_{0})$, where $a_{0}$ is the
constant MOND acceleration and $\mu$ the interpolating function.
The structure of the second term depends greatly
on the angle between $\bmath{g}_{\rm ext}$ and
and the velocity of the perturber. We evaluate the dynamical drag force
numerically and compare our MOND results with the Newtonian case.
We mention the relevance of our calculations to orbit evolution
of globular clusters and satellites
in a gaseous proto-galaxy. Potential differences
in the X-ray emission of gravitational galactic wakes
in MOND and in Newtonian gravity with a dark halo are highlighted. 
\end{abstract}


\section{Introduction}
According to the Brownian theory,
a `macroscopic' particle moving in a fluid experiences
a fluctuating force as a consequence of the graininess of the fluid,
resulting in dynamical friction (DF) and diffusion in the velocity-space.
Understanding the nature of the DF force experienced
by a gravitational object that moves against a mass
density background is of great importance for describing the evolution of
gravitational systems and the exchange of angular and linear momentum
between gravitational subsystems.
There are two limits of interest in
astrophysics: pure collisionless media and fully collisional gas. 
If the pertuber only interacts with the surrounding particles gravitationally,
either the medium is collisionless or collisional,
the DF force is the result of the formation of 
an overdensity wake behind the perturber. 

The DF depends on the nature of the force between
the perturber and the background particles. 
Departures from Newtonian gravity have been proposed as an alternative
to dark matter in galaxies (e.g., Modified Newtonian Dynamics;
Milgrom 1983).  
Ciotti \& Binney (2004) found that in collisionless backgrounds,
the characteristic timescale for 
DF in Modified Newtonian Dynamics (MOND) is much shorter than in
the Newtonian equivalent system with dark matter. If confirmed,
this result may have strong astrophysical
implications and may be used to distinguish between dark matter
and modified gravity. Using the MOND scaling found in Ciotti \& Binney (2004), 
the globular clusters in Fornax
dwarf spheroidal galaxy should have spiraled into the centre of Fornax
in $\sim 0.5 $ Gyr, forming a visible galactic nucleus
(S\'anchez-Salcedo et al.~2006). 

Ciotti \& Binney (2004) derived the MOND DF timescale in a 
restrictive plane-parallel geometry, which is a too
specialized case, and in a non-standard
statistical formulation. 
These caveats led Nipoti et al.~(2008) to carry out fully N-body
simulations of the dynamics of bars in MOND systems. Their
simulations confirmed the scaling of the DF timescale predicted
in Ciotti \& Binney (2004). Still, it is not entirely clear  
how the background responds to the perturbation, and the
description of the 
structure of the density wake induced by a small and massive perturber
is lacking. 
Since MOND is a nonlinear theory, the gravitational acceleration
induced by the perturber is affected by the presence of an external 
gravitational field. In particular,  
the DF force is expected to depend
on the angle between the direction of the mean field and the
velocity of the perturber. Our aim is to understand the nature
of DF in MOND. 

In this paper we consider gaseous
media and study the structure of the density wake induced by
a gravitational perturber in MOND.  Gaseous DF
has found many applications in astrophysics, ranging from protoplanet accretion all the way to the motion of galaxies in clusters 
(e.g., Ostriker 1999;
Kim 2007; Conroy \& Ostriker 2008). In a pure baryonic universe as in MOND
theory, the relative role of gas becomes even more important
because all the contribution to the mass is accounted for by the
gas and the stars. 
Moreover, the complementary view of hydrodynamics may provide new insights
into the differences and analogies between
Newtonian and modified gravities.
Although the details on the internal density structure of the induced wake
may depend on whether the system is fully collisional or collisionless,
the fluid limit is useful for understanding
the main conceptual features introduced by the 
change of the gravity law.

The linear response of the gaseous medium to a gravitational perturber
in Newtonian gravity is well-documented (Dokuchaev 1964;
Ruderman \& Spiegel 1971; Just \& Kegel 1990;
Ostriker 1999; S\'anchez-Salcedo \& Brandenburg 1999, 2001;
Kim \& Kim 2007; Kim et al.~2008).
In all these works, a minimum radius $r_{\rm min}$ in the Coulomb
logarithm was introduced in order to regularise the gravitational 
potential of a point mass.
Although the formulae were derived for rectilinear orbits in homogeneous
and infinity media, simple `local' extensions have been proven very
successful in more realistic situations, e.g., smoothly decaying density
backgrounds or when the perturber is moving on a circular orbit 
(S\'anchez-Salcedo \& Brandenburg 2001; Kim \& Kim 2007).  
 
The paper is organized as follows. 
In \S 2, we discuss the basic concepts on 
the ideal problem of a point particle moving at constant speed
within a uniform gas  
in MOND (the Bondi-Hoyle problem). In \S 3, we outline the linear
derivation of the basic equations for calculating the steady-state
density wake generated by an extended body.  In \S 4 we 
describe the structure of the resulting wake and 
evaluate the DF force exerted on the perturber. 
We then discuss some implications
of our results in \S 5. Finally, we conclude in \S 6.

\section{The Bondy-Hoyle problem in MOND}
We consider a gravitational point particle at the origin of our 
coordinate system, 
surrounded by a gas whose velocity far from the particle is
\begin{equation}
\bmath{v}_{\infty}={\mathcal{M}}c_{\infty}\hat{z},
\label{eq:vz}
\end{equation}
where $c_{\infty}$ is the sound speed of the gas at infinity and
${\mathcal{M}}$ is the Mach number. 
We are interested in the wake produced by the gravitational interaction
with the perturber in the context of MOND. 

Bekenstein \& Milgrom (1984) suggested a Lagrangian theory and also a 
nonlinear differential equation for the nonrelativistic MOND gravitational 
potential produced by a mass density distribution $\rho$:
\begin{equation}
\bmath{\nabla}\cdot\left[\mu\left(\frac{|\bmath{\nabla}\Phi|}{a_{0}}\right)
\bmath{\nabla}\Phi\right]=4\pi G\rho,
\label{eq:poissonMOND}
\end{equation}
where $a_{0}\sim 10^{-8}$ cm s$^{-2}$ is an universal acceleration 
and $\mu(x)$ is a monotonic
and continuous function with
the property that $\mu(x)\simeq x$ for $x\ll 1$ (deep MOND regime)
and $\mu(x)\simeq 1$ for $x\gg 1$ (Newtonian regime).
Only for very special configurations  (one-dimensional symmetry --spherical,
cylindrical or plane symmetric systems-- or Kuzmin discs)
the MOND acceleration $\bmath{g}$ is related to the Newtonian 
acceleration, $\bmath{g}_{N}$ by the algebraic
relation $ \mu(|\bmath{g}|/a_{0})\bmath{g}=\bmath{g}_{N}$ 
(Brada \& Milgrom 1995).
At variance with the Poisson equation, MOND is a non-linear theory. 
This implies that the gravitational
potential generated by the perturber 
depends in a complex way on the external field
$\bmath{g}_{\rm ext}$ in which it is immersed. 
The external field is to be thought of as the field of an enveloping
system in the absence of the perturber.
For example, the perturber can be a galaxy in the external
field of a cluster of galaxies, or a dwarf spheroidal galaxy in
the field of its parent galaxy. In MOND, it is crucial 
to include the external field $\bmath{g}_{\rm ext}$ to describe
DF on a perturber. The case $\bmath{g}_{\rm ext}=0$ is not 
very relevant when considering DF because it corresponds
to a situation where all the
surrounding gas is bound to the perturber, whereas
DF concerns the interaction with {\it unbound} distant particles having 
large impact parameters. MOND permits to have unbound particles 
if $\bmath{g}_{\rm ext}\neq 0$.
For these reasons, we will assume that $\bmath{g}_{\rm ext}$ is a 
no-null vector field.

Along this paper we will assume that the unperturbed gas
density is constant all over the space and that the body moves in a 
uniform rectilinear orbit,  
which seems to be in contradiction with the inherent 
assumption that there is an external gravitational field.
Note, however, that this approximation is amply used even in Newtonian
dynamics where in most of the cases (e.g., a galaxy inside the galaxy cluster) 
the gravitational body is immersed in a stratified medium. 
The `local' approximation, that is, estimating the drag force
at the present location of the perturber as if the medium
were homogeneous but taking appropriately the Coulomb logarithm,
has been proven very successful in both collisionless and collisional
systems (e.g., S\'anchez-Salcedo \& Brandenburg 2001, and references therein).
So far we are not interested in including the complications
of density gradients in the unperturbed surrounding medium.

\subsection{The far-field equation: external dominated field  }
Far away enough from the object, the change in the external gravitational
potential can be treated as a linear perturbation.  This is valid
as soon as the physical size of the perturber is supposed to
be small as compared to the characteristic length-scale of the medium.
Using the subscript $0$ for referring to unperturbed quantities,
the linear field equation is
\begin{equation}
\bmath{\nabla}\cdot \left[\mu_{0}\left(1+
L_{0}\tilde{\mathcal{E}}_{0}\right)\cdot\bmath{\nabla}\Phi_{1}\right]
=4\pi G\rho_{1},
\label{eq:lineal}
\end{equation}
where $\Phi_{1}$ is the change in $\Phi$ produced by an increment 
$\rho_{1}$ in $\rho$ (Milgrom 1986). Here $\mu_{0}\equiv\mu(g_{\rm ext}/a_{0})$,
$L_{0}$ is the logarithmic derivative
of $\mu_{0}$ (in the unperturbed system) and 
$\tilde{\mathcal{E}}_{0}$ represents a $3\times 3$ matrix with
elements $\tilde{\mathcal{E}}_{0,i,j}=\hat{\bmath{e}}_{0,i}
\hat{\bmath{e}}_{0,j}$, where 
$\hat{\bmath{e}}_{0}$ is the unit vector in the direction
of $\bmath{g}_{\rm ext}$, which is $\bmath{r}$ dependent. 
As Milgrom (1986) pointed out, $\Phi_{1}$
satisfies an equation analogous to the electrostatic field equation
with an inhomogeneous and anisotropic dielectric tensor.

Since we are not interested in the distortions in the wake
by tidal forces,
we will assume that $\bmath{g}_{\rm ext}$ is a constant vector all over
the space. This assumption also facilitates a comparison with the
more familiar Newtonian case.  If so,
$\mu_{0}$, $L_{0}$ and $\hat{\bmath{e}}_{0}$ are also constant
and the linear field equation for a constant external field can be written
as (Milgrom 1986):
\begin{equation}
\nabla^{2}\Phi_{1}+L_{0}\frac{\partial^{2}\Phi_{1}}{\partial z^{2}}=
4\pi \mu_{0}^{-1}G\rho_{1}.
\label{eq:m86}
\end{equation}
It can be seen that the potential becomes Newtonian (but
with a larger effective gravitational constant) and anisotropic.
We must note that, although the above equation was derived 
for $|\bmath{\nabla}\Phi_{1}|
\ll g_{\rm ext}$, it
is also valid when $g_{\rm ext}\gg a_{0}$, regardless the value
of $|\bmath{\nabla}\Phi_{1}|$. In this limit
$\mu_{0}\approx 1$ and $L_{0}\approx 0$; the conventional Poisson
equation is recovered.

\subsection{The Bondi-Hoyle radius}
\label{sec:bondi}

In the classical problem of a slow Brownian particle in a fluid,
the field particles are assumed to form a heat bath.
That is, they all stay close to thermal equilibrium, despite the presence
of the Brownian particle and, hence, the equation of motion is
solved by perturbation. 
In the case of a point gravitational perturber 
immersed in a perfect gaseous medium, the Bondi-Hoyle radius
defines the region where the response of the gas is linear.
In Newtonian dynamics the Bondi-Hoyle radius is 
$r_{BH}\equiv GM/(c_{\infty}^{2}(1+{\mathcal{M}}^{2}))$.
Streamlines whose impact parameter is less than $2r_{BH}$, 
will bend significantly and pass through a shock.
Hence, within $2r_{BH}$ it is not any longer a small perturbation.
In order to regularise the gravitational potential of a point mass,
one has to introduce a minimum radius in the formulae of the
DF drag ($r_{\rm min}\approx 2r_{BH}$).

In the Appendix \ref{sec:appBondi}, we estimate the Bondi-Hoyle 
radius for a point mass in MOND.
It is shown that
the MOND Bondi-Hoyle radius is larger than in Newtonian gravity
by a factor between 
$\mu_{0}^{-1}$ and $\mu_{0}^{-1}(1+L_{0})^{-1/2}$,
depending on the angle between the velocity of the particle
and the external field.
To get a sense of values of $r_{BH}$ in typical cases, consider a
galaxy of $5\times 10^{11}$
M$_{\odot}$ orbiting supersonically ${\mathcal{M}}\approx 1.5$
in a cluster of galaxies with a sound speed of intracluster gas
of $\sim 1500$ km s$^{-1}$. 
The Bondi-Hoyle radius in this case is $r_{BH}\lesssim \mu_{0}^{-1}$ kpc.
In a typical galactic cluster $\mu_{0}=0.3$--$1$ (see, e.g., fig.~7
in Sanders \& McGaugh 2002), hence $r_{BH}\lesssim 1$--$3$ kpc,
implying that, if the interaction with the
intracluster gas is merely gravitational, the linear approximation
is satisfactory for studying the gaseous wake even quite close to
the galaxy. 
For extended perturbers with characteristic size
much larger than $2r_{BH}$, the 
flow is essentially laminar at any location.
In the remainder of the paper we will describe the perturbation
on the gas using linear theory.

\section{Formulation}
\label{sec:lineareq}
\subsection{Modelling the perturber}
Our aim is to study the large-scale gravitational perturbation 
induced by a small 
perturber travelling through a much larger system.
We obtain that, beyond a certain distance from the
perturber, $|\bmath{\nabla}\Phi_{1}|\ll g_{\rm ext}$.
In other words, the far-field wake is expected to be in the external
field-dominated regime.
In order to highlight the differences between genuine MOND and Newtonian
gravity, we restrict our considerations to situations in which
the potential is dominated by the external field everywhere,
i.e., we 
will consider an extended perturber of mass $M$ and characteristic
size $r_{p}$, 
where the internal acceleration $g_{\rm int}\approx 
GM_{p}/(\mu_{0}r_{p}^{2})\ll g_{\rm ext}$\footnote{ 
For a point-like perturber, one can always find a vicinity
of the body where the inequality $g_{\rm int}\ll g_{\rm ext}$
is not achieved.}.
The following density-potential pair, which
corresponds to a ``modified'' Plummer model,
is an exact solution of Eq.~(\ref{eq:m86})
\begin{equation}
\rho_{p}(\bmath{r})=\frac{3Mr_{p}^{2}}{4\pi\sqrt{1+L_{0}}\left(x^{2}+y^{2}+
\frac{z^{2}}{1+L_{0}}+r_{p}^{2}\right)^{5/2}},
\label{eq:densityaxi}
\end{equation}
\begin{equation}
\Phi_{p}(\bmath{r})=-\frac{1}{\mu_{0} }
\frac{GM}{\sqrt{(1+L_{0})(x^{2}+y^{2})+z^{2}+\bar{r}_{p}^{2}}},
\label{eq:potentialaxi}
\end{equation}
where $r_{p}$ and $\bar{r}_{p}\equiv (1+L_{0})^{1/2} r_{p}$ are the 
characteristic radii.  
Note that for $L_{0}=0$ and $\mu_{0}=1$ (Newtonian limit), it corresponds
to the classical spherical Plummer model.
Sometimes it is useful to express $\rho_{p}$ in terms of the 
central density of the object, $\rho_{c}$, as follows 
\begin{equation}
\rho_{p}(\bmath{r})=\rho_{c}\left[1+\frac{x^{2}}{r_{p}^{2}}
+\frac{y^{2}}{r_{p}^{2}}+
\frac{z^{2}}{(1+L_{0})r_{p}^{2}}\right]^{-5/2},
\end{equation}
where
\begin{equation}
\rho_{c}=\frac{3M}{4\pi (1+L_{0})^{1/2} r_{p}^{3}}.
\end{equation}
The central density can be written in terms of the central
density for the spherical Plummer model in classical Newtonian
gravity, $\tilde{\rho}_{c}$, as $\rho_{c}=\tilde{\rho}_{c}/(1+L_{0})^{1/2}$.

The selection of a Plummer model was for analytical purposes.
As long as the size of the system is much larger
than the perturber, the structure of the wake and the drag force 
experienced by the
perturber are not expected to be sensitive to the details of the potential
close to the body. 

In order to approach this problem analytically, we study the simplest
case: an externally dominated perturber. In real life,  
there are 
some Galactic dwarf spheroidal galaxies that are known to be in this
regime (e.g., Milgrom 1995; S\'anchez-Salcedo \& Hernandez 2007). It is
likely that some subclusters and groups of galaxies, 
with low internal accelerations, 
embedded in a main massive galaxy cluster (such as the NGC 4911 group in the
Coma Cluster) lie also in this regime (e.g., Sanders \& McGaugh 2002).

In Eqs (\ref{eq:densityaxi}) and (\ref{eq:potentialaxi}),  
the external field was taken along $z$ and, therefore, in the same
direction as the incident flow (see Eq.~\ref{eq:vz}). In this section
we will focus on this {\it axisymmetric case}. 
The derivation of the equations
when $\bmath{g}_{\rm ext}$ is perpendicular 
to $\bmath{v}_{\infty}$ is postponed up to
\S \ref{sec:orthogonalsec}.
These two situations brackets a general
case where the external field has an arbitrary angle with respect
to the velocity of the flow at infinity.

\subsection{Linear equations in an external dominated field}
\label{sec:equationswake}
In the following, we give
the linear derivation of the wake
in a medium with unperturbed density
$\rho_{0}$ and adiabatic sound speed $c_{\infty}$, 
ignoring gas self-gravity and any magnetic fields. 
As stated in Eq.~(\ref{eq:vz}),
it is assumed that the gravitational perturber is 
seated at the origin of our coordinate system and the
gas velocity far from the perturber is
$\bmath{v}_{\infty}={\mathcal{M}}c_{\infty}\hat{z}$.
In the axisymmetric case,
$\bmath{v}_{\infty}$ and $\bmath{g}_{\rm ext}$ are parallel.
We are interested in the steady-state density enhancement
${\mathcal{D}}(\bmath{r})=(\rho-\rho_{0})/\rho_{0}$
produced by the gravitational interaction with the perturber. 
The steady-state linearized basic dynamical equations for 
adiabatic perturbations 
$\rho=\rho_{0}+\rho'$ and $\bmath{v}=\bmath{v}_{\infty}+ \bmath{v}'$ are:
\begin{equation}
\rho_{0}\bmath{\nabla}\bmath{v}'+\bmath{v}_{\infty}\cdot\bmath{\nabla}\rho'=0,
\label{eq:continuity}
\end{equation}
and
\begin{equation}
(\bmath{v}_{\infty}\cdot\bmath{\nabla})\bmath{v}'=
-\frac{c_{\infty}^{2}}{\rho_{0}} \bmath{\nabla}\rho'-\bmath{\nabla}\Phi_{p}.
\label{eq:motion}
\end{equation}
Our strategy is to eliminate $\bmath{v}'$ everywhere.
By substituting equation (\ref{eq:continuity}) in the divergence
of equation (\ref{eq:motion}), we obtain that 
${\mathcal{D}}$ satisfies the differential equation
\begin{equation}
L{\mathcal{D}}=-\frac{1}{c_{\infty}^{2}}\nabla^{2}\Phi_{p},
\label{eq:ld}
\end{equation}
where $L$ is the linear differential operator
\begin{equation}
L{\mathcal{D}}\equiv
\frac{\partial^{2}{\mathcal{D}}}{\partial x^{2}}+
\frac{\partial^{2}{\mathcal{D}}}{\partial y^{2}}+
\left(1-{\mathcal{M}}^{2}\right)\frac{\partial^{2}{\mathcal{D}}}{\partial z^{2}}.
\label{eq:operatorL}
\end{equation}
The operator $L$ arises frequently in fluid dynamics (e.g., Landau \&
Lifshitz 1959). 

In Eqs~(\ref{eq:continuity})-(\ref{eq:operatorL})  we have not specified 
the law of gravity.
In MOND, the Laplacian of the potential in Eq.~(\ref{eq:ld}) can be 
expressed in terms of $\rho_{p}$ using the MOND field equation (\ref{eq:m86}).
The equation for ${\mathcal{D}}$ becomes 
\begin{equation}
L{\mathcal{D}}
=-\frac{4\pi G \rho_{p}}{\mu_{0}c_{\infty}^{2}}
+\frac{L_{0}}{c_{\infty}^{2}}\frac{\partial^{2}\Phi_{p}}{\partial z^{2}}.
\label{eq:phizz}
\end{equation}

From now on, it will be convenient to use the following dimensionless variables:
$\hat{x}=x/r_{p}$, $\hat{y}=y/r_{p}$
and $\hat{z}=z/(1+L_{0})^{1/2}r_{p}$. The hat symbol over a certain
variable $\chi$ will be used to denote that $\chi$ is written with 
$\hat{x}$, $\hat{y}$ and $\hat{z}$ as the 
arguments\footnote{At this
stage it is obvious that this transformation does not conserve
mass in the sense that if $\rho$ is a density field, then
$\int \rho dx dy dz\neq \int\hat{\rho}d\hat{x}d\hat{y}d\hat{z}$.}, 
e.g., $\hat{D}(\hat{x},\hat{y},\hat{z})=D(x,y,z)$.
The second-order derivative in Eq.~(\ref{eq:phizz})
can be performed as soon as the potential is known.
Evaluating the second-order derivative of the potential
in Eq.~(\ref{eq:phizz})
using the potential given in Eq.~(\ref{eq:potentialaxi}),
and rearranging and grouping the terms,
the solution $\hat{\mathcal{D}}$
can be expressed as a linear superposition of two contributions 
$\hat{\mathcal{D}}=\hat{\mathcal{D}}_{1}+\hat{\mathcal{D}}_{2}$. Each
one satisfies the following differential equations:
\begin{equation}
\hat{L}{\hat{\mathcal{D}}}_{1}=
-\left[1-T_{0}\right] \frac{4\pi G\rho_{c}r_{p}^{2}}{\mu_{0}c_{\infty}^{2}}
\hat{g}_{1}(\hat{\bmath{r}}),
\label{eq:master1}
\end{equation}
and
\begin{equation}
\hat{L}{\hat{\mathcal{D}}}_{2}=
-T_{0}\frac{4\pi G\rho_{c}r_{p}^{2}}{\mu_{0}c_{\infty}^{2}}
\hat{g}_{2}(\hat{\bmath{r}}),
\label{eq:master2}
\end{equation}
with
\begin{equation}
\hat{g}_{1}(\hat{\bmath{r}})=\frac{1}{[1+\hat{r}^{2}]^{5/2}},
\end{equation}
\begin{equation}
\hat{g}_{2}(\hat{\bmath{r}})=
\frac{2\hat{z}^{2}-\hat{x}^{2}-\hat{y}^{2}}{[1+\hat{r}^{2}]^{5/2}},
\end{equation}
where $\hat{r}^{2}=\hat{x}^{2}+
\hat{y}^{2}+\hat{z}^{2}$,
\begin{equation}
T_{0}=\frac{L_{0}}{3(1+L_{0})},
\label{eq:todef}
\end{equation} and the operator $\hat{L}$ is
\begin{equation}
\hat{L}=
\frac{\partial^{2}}{\partial \hat{x}^{2}}+
\frac{\partial^{2}}{\partial \hat{y}^{2}}+
\left(1-{\mathcal{M}}^{2}_{\rm eff}\right)\frac{\partial^{2}}{\partial \hat{z}^{2}},
\end{equation}
with 
\begin{equation}
{\mathcal{M}}^{2}_{\rm eff}\equiv \frac{{\mathcal{M}}^{2}+L_{0}}{1+L_{0}}.
\label{eq:meff}
\end{equation}
According to Eq.~(\ref{eq:todef}), 
$T_{0}$ varies from $0$ in the Newtonian regime ($L_{0}=0$)
to $1/6$ in the deep MOND regime (i.e.~$L_{0}=1$).
From Eq.~(\ref{eq:meff}) it can be seen that 
${\mathcal{M}}_{\rm eff}\geq {\mathcal{M}}$ 
in the subsonic case, 
${\mathcal{M}}_{\rm eff}\leq {\mathcal{M}}$ 
in the supersonic case, and 
${\mathcal{M}}_{\rm eff}={\mathcal{M}}=1$ 
at the transonic velocity.

We recall that the equation for $\hat{\mathcal{D}}$ in the Newtonian case
is
\begin{equation}
\hat{L}\hat{\mathcal{D}}
=-\frac{4\pi G\tilde{\rho}_{c}r_{p}^{2}}{c_{\infty}^{2}}g_{1}(\hat{\bmath{r}}),
\label{eq:newtoniancase}
\end{equation}
which is naturally recovered from the above equations just
by taking $\mu_{0}=1$ and $L_{0}=0$ (so that ${\mathcal{M}}_{\rm eff}=
{\mathcal{M}}$).
The component $\hat{\mathcal{D}}_{1}$ obeys a differential equation 
similar to the Newtonian case and hence is identical
to the wake induced by a perturber with mass density 
$\hat{\rho}_{1}=(1-T_{0})(1+L_{0})^{-1/2}\tilde{\rho}_{c}g_{1}
(\hat{\bmath{r}})$ 
in conventional Newtonian gravity,
once $G$ has been replaced by a larger effective value $G/\mu_{0}$.
The profile $\hat{\rho}_{1}$ corresponds to the classical 
(spherical) Plummer model in 
the transformed coordinates ($\hat{x},\hat{y},\hat{z}$),
multiplied by a form factor between $0.59$ (deep-MOND limit)
and $1$ (Newtonian limit). 
In analogy to the Newtonian case (Eq.~\ref{eq:newtoniancase}), 
a fictitious Newtonian
perturber with a pseudo-density mass density distribution
 $\hat{\rho}_{2}=T_{0}\rho_{c}\hat{g}_{2}$ 
generates a component identical to $\hat{\mathcal{D}}_{2}$. 
We refer to $\hat{\rho}_{2}$ as pseudo-density because it may take negative 
values. Interestingly, the total
mass associated with this distribution is
\begin{equation}
\hat{M}_{2}={\mathcal{PV}}\int_{-\infty}^{\infty}\int_{-\infty}^{\infty}
\int_{-\infty}^{\infty} \hat{\rho}_{2} \;d\hat{x} d\hat{y} d\hat{z}=0.
\end{equation}
Our natural choice was to adopt the Cauchy principal value of the
integral, which we denote by the symbol ${\mathcal{PV}}$
(e.g., Mathews \& Walker 1970).
Note that $\hat{\rho}_{2}$ decays more slowly with radius
than $\hat{\rho}_{1}$. Hence, the pseudo-density $\hat{\rho}_{2}$
becomes larger than $\hat{\rho}_{1}$ at large radii.
The importance of the contribution of ${\mathcal{D}}_{2}$ to the wake
and to the gravitational drag is
difficult to forsee without a quantitative study.

\subsection{Formal solution}
The solution of Eqs~(\ref{eq:master1}) and (\ref{eq:master2})
can be obtained using the retarded
Green's function, which may be derived following different paths
(e.g., Just \& Kegel 1990; Ostriker 1999). In particular, the
Green function can be found after Fourier transforming
and imposing causality when choosing the contour of integration in 
the complex plane for ${\mathcal{M}}_{\rm eff}>1$ (e.g., Just \& Kegel
1990; Furlanetto \& Loeb 2002). 
In the steady-state, the perturbed density fields $\hat{\mathcal{D}}_{1}$
and $\hat{\mathcal{D}}_{2}$ are
\begin{equation}
\hat{\mathcal{D}}_{1}=\frac{1-T_{0}}{(1+L_{0})^{1/2}\mu_{0}}
\frac{G\tilde{\rho}_{c}r_{p}^{2}}
{c_{\infty}^{2}}\hat{I}_{1},
\end{equation} 
\begin{equation}
\hat{\mathcal{D}}_{2}=\frac{T_{0}}{(1+L_{0})^{1/2}\mu_{0}}
\frac{G\tilde{\rho}_{c}r_{p}^{2}}
{c_{\infty}^{2}}\hat{I}_{2},
\label{eq:master2b}
\end{equation} 
where
\begin{equation}
\hat{I}_{i}(\hat{\bmath{r}})=
\int d^{3}\hat{\bmath{r}}' \frac{\xi' \hat{g}_{i}(\hat{\bmath{r}}')}
{[(\hat{z}-\hat{z}')^{2}-\beta_{\rm eff}^{2}
(\hat{R}^{2}+\hat{R}'^{2}-2\hat{R} \hat{R}'\cos\theta')]^{1/2}},
\label{eq:master2c}
\end{equation} 
for $i=1,2$, with $\beta^{2}_{\rm eff}\equiv {\mathcal{M}}_{\rm eff}^{2}-1$ and
\[ \xi' = \left\{ \begin{array}{ll}
         2 & \mbox{if ${\mathcal{M}}_{\rm eff}>1$, and}\\
           & \mbox{$z-z'+\beta_{\rm eff}
[R^{2}+R'^{2}-2RR'\cos\theta']^{1/2}>0$;}\\
         1 & \mbox{if ${\mathcal{M}}_{\rm eff}<1$;}\\
         0 & \mbox{otherwise.} \end{array} \right. \] 

\section{Results}
\subsection{The density structure of the wake}
\subsubsection{The component ${\mathcal{D}}_{1}$ in the axisymmetric case}
\begin{figure}
\epsfig{file=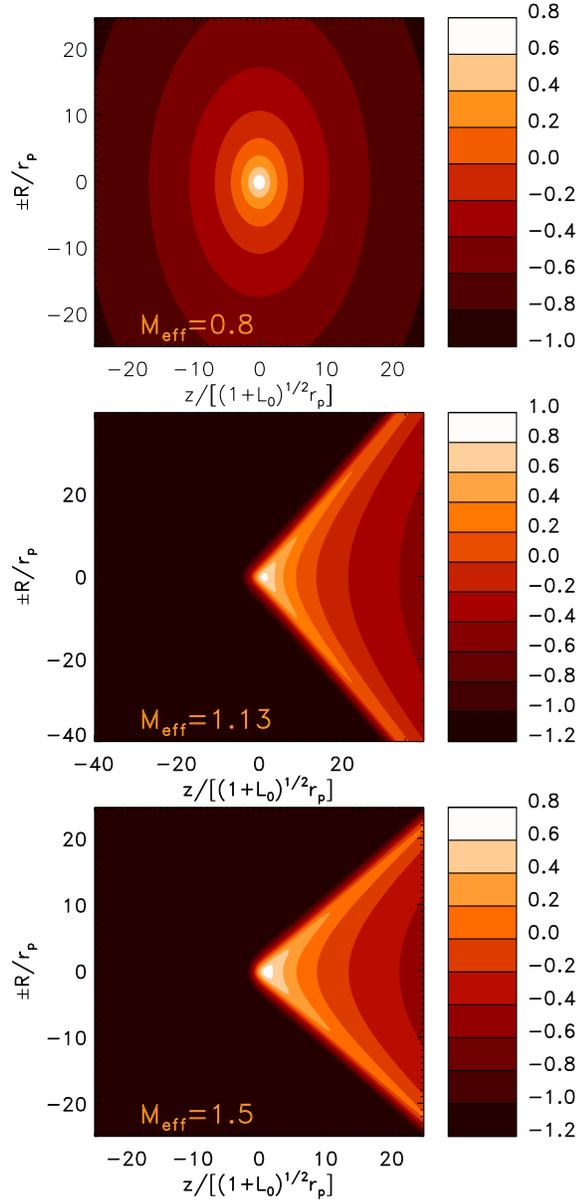,angle=0,width=17.5cm}
  \caption{Distribution of the dimensionless integral $\hat{I}_{1}$, in
logarithmic scale, for ${\mathcal{M}}_{\rm eff}=0.8, 1.13$ and
$1.5$ in the axisymmetric case. The box size for 
${\mathcal{M}}_{\rm eff}=1.13$ is larger in order to have the same dynamical
range in density.}
\label{fig:D1axi}
\end{figure}

Figure \ref{fig:D1axi} shows the integral $\hat{I}_{1}$, 
which is proportional to
the perturbed density $\hat{D}_{1}$, in the $\hat{z}$--$\hat{R}$ plane,
of a ${\mathcal{M}}_{\rm eff}=0.8, 1.13, 1.5$ body. 
In the deep MOND regime, they correspond to physical
Mach numbers of $0.53$, $1.25$ and $1.9$, respectively.
So far we are only interested in the `far-field' perturbed density, 
hence we will not delve into details regarding the near-field (within a few 
core radius from the perturber). A subsonic perturber generates
a density distribution with contours of constant density corresponding
to similar ellipses with eccentricity ${\mathcal{M}}_{\rm eff}$.
For supersonic motions, however, the region
of perturbed density is confined within the rear Mach cone,
dragged by its apex by the perturber. The surfaces of constant
density within the wake correspond to hyperbolae in the 
$\hat{z}$--$\hat{R}$ plane,
with eccentricity $e={\mathcal{M}}_{\rm eff}$. This is expected
because, as we
show in Section \ref{sec:equationswake}, the equation for 
$\hat{\mathcal{D}}_{1}$ has the same form as in the 
Newtonian case with ${\mathcal{M}}_{\rm eff}$,
once the density distribution of the perturber is
rescaled by a factor $(1-T_{0})(1+L_{0})^{-1/2}$, and $G$ is replaced by
$G/\mu_{0}$. Using this analogy, we can take advantage of
the analytical results of Ostriker (1999) to find the 'far-field'
perturbed density:
\begin{eqnarray}
{\mathcal{D}}_{1}&\simeq &\frac{1-T_{0}}{(1+L_{0})^{1/2}}
\frac{\mu_{0}^{-1}\xi'' G\hat{M}_{1}r_{p}^{2}/c_{\infty}^{2}}
{\sqrt{\hat{z}^{2}-\beta_{\rm eff}^{2}
\hat{R}^{2}}}\\
&=& (1-T_{0})
\frac{\mu_{0}^{-1}
\xi'' GM/c_{\infty}^{2}}{\sqrt{z^{2}-
\beta^{2}R^{2}}},
\label{eqn:isoconA}
\end{eqnarray}
where $\xi''=2$ for supersonic perturbers and $1$ in the subsonic regime.
In Eq.~(\ref{eqn:isoconA}) we used that 
$\hat{M}_{1}=\tilde{\rho}_{c}\int \hat{g}_{1} d\hat{x}d\hat{y}
d\hat{z}=M/r_{p}^{3}$.
From the equation above, we see that
the isodensity contours are $z^{2}+R^{2}(1-{\mathcal{M}}^{2})=$const,
i.e. ellipses or hyperbolae with
eccentricity $e={\mathcal{M}}$.
In the Appendix \ref{sec:appendixB}
we reconsider the perturbation as a time-dependent rather than
a steady state problem. 

\subsubsection{The component ${\mathcal{D}}_{2}$ in the axisymmetric case}
\begin{figure}
\epsfig{file=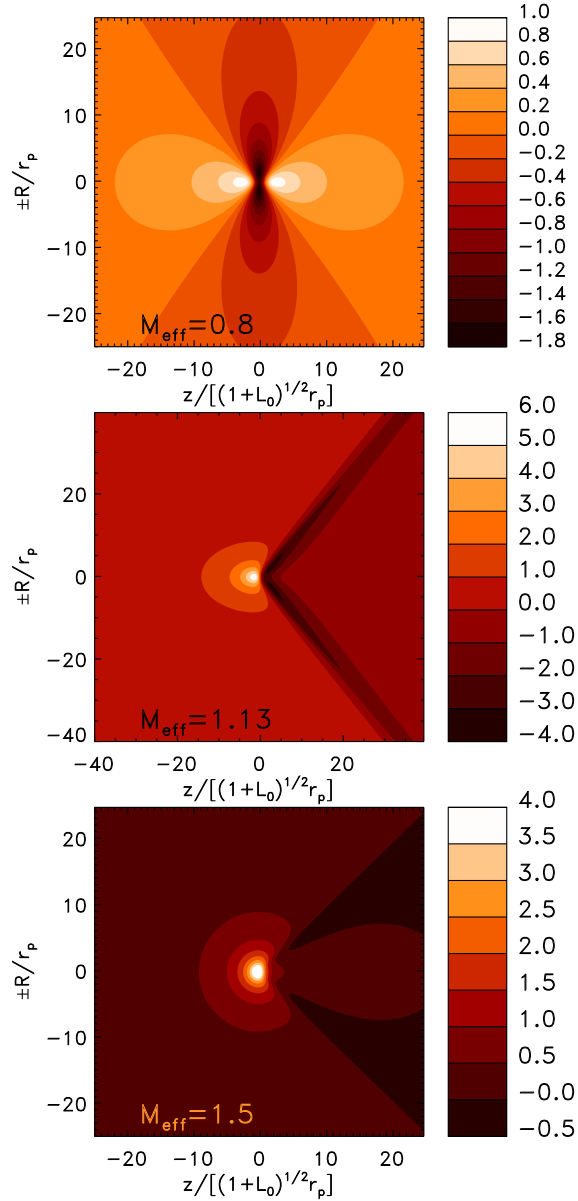,angle=0,width=17.5cm}
  \caption{Profile of the dimensionless integral $\hat{I}_{2}$,
in the axisymmetric case.
The scale is linear scale. 
The effective Mach numbers are the same as in
Fig.~\ref{fig:D1axi}.
}
\label{fig:I2axi}
\end{figure}

Figure \ref{fig:I2axi} shows $\hat{I}_{2}$ for the same three 
effective Mach numbers as previously considered 
(${\mathcal{M}}_{\rm eff}=0.8$, $1.13$ and $1.5$).
In the subsonic regime, $\hat{I}_{2}$, with a bipolar structure,
is positive along
the $z$ axis and negative in the perpendicular plane.
For supersonic perturbers, an overdense bump at the head of
the perturber is generated.
Interestingly, $\hat{I}_{2}$ displays a drop in density along
the surface of the Mach cone.
This negative jump in the surface of the Mach cone is very remarkable at 
${\mathcal{M}}_{\rm eff}=1.13$ and dilutes at larger
effective Mach numbers. Our calculations show that at
${\mathcal{M}}_{\rm eff}=1.75$ there is no
jump in the Mach cone surface.
For perturbers with ${\mathcal{M}}_{\rm eff}>1.75$ the density
jump in the Mach cone becomes positive.
We see that, in general, $\hat{I}_{2}$ may take values comparable
to $\hat{I}_{1}$. We note, however, that
the real density $\hat{D}_{2}$ is related to $\hat{I}_{2}$ through a 
factor $T_{0}\leq 1/6$ (see Eq.~\ref{eq:master2b}).

\begin{figure}
\epsfig{file=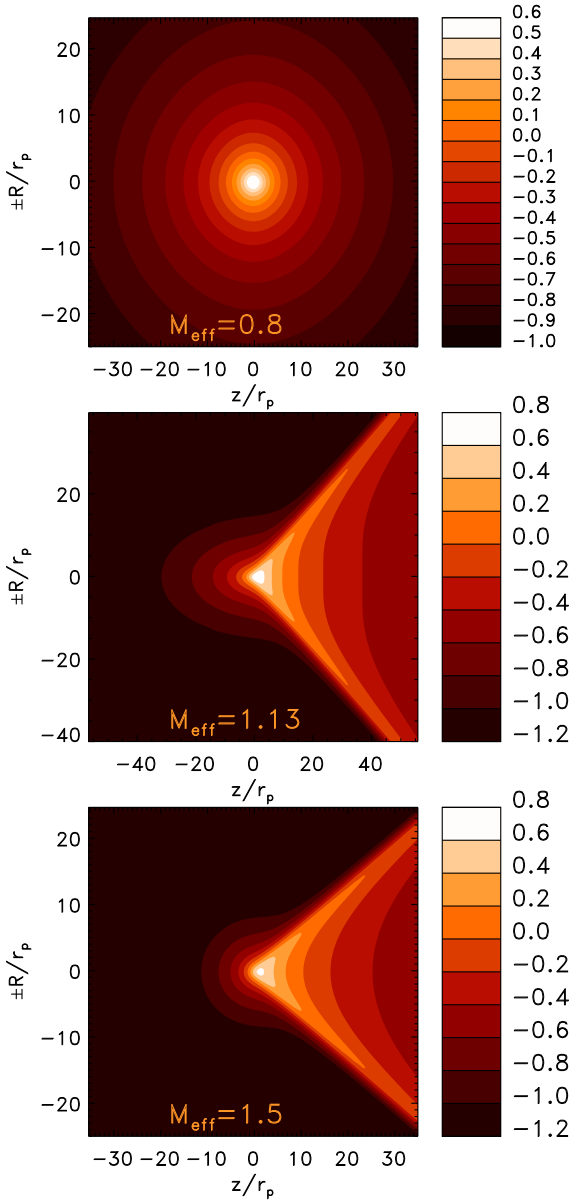,angle=0,width=17.5cm}
  \caption{Density perturbation profiles, $\rho/\rho_{0}$, in units of
$\mu_{0}^{-1}[G\tilde{\rho}_{c}r_{p}^{2}/\sqrt{2}c_{\infty}^{2}]$
in the axisymmetric case and with $L_{0}=1$. The scale is logarithmic. }
\label{fig:superpositionaxi}
\end{figure}

Figure \ref{fig:superpositionaxi} contains the superposition 
$(1-T_{0})\hat{I}_{1}+T_{0}\hat{I}_{2}$, which is proportional
to ${\mathcal{D}}$, in
the deep MOND regime (i.e.~$L_{0}=1$ and $T_{0}=1/6$). 
In the subsonic regime, the inclusion of the component ${\mathcal{D}}_{2}$
plays an important role. 
The resulting contours of isodensity can be fitted by ellipsoids
defined by the equation
$R^{2}+z^{2}/q^{2}=$const, where $q$ is the flattening parameter
of the distribution in the {\it physical} $R$--$z$ plane.
According to our discussion in the preceding section, for
${\mathcal{M}}_{\rm eff}=0.8$ and $L_{0}=1$ (i.e.~${\mathcal{M}}=0.53$), 
we know that the isodensity contours of ${\mathcal{D}}_{1}$ have $q=0.85$.
The superposition of components ${\mathcal{D}}_{1}$ plus
${\mathcal{D}}_{2}$
generates a density distribution with $q=1.23$ (in the deep-MOND
regime).
In order to have an axis ratio of $q=0.85$, the body
should be moving at ${\mathcal{M}}=0.85$. 
For ${\mathcal{M}}=0.2$, the flattening parameter is $0.98$ if
only ${\mathcal{D}}_{1}$ is considered, and becomes $1.38$ when 
${\mathcal{D}}_{2}$ is
also taken into account. 
As a general conclusion, the wake for a subsonic body
is more flattened along the direction of motion in MOND than
in Newtonian dynamics.

In the supersonic deep-MOND case, the overdensity head is still visible
when both components are added. The inclusion
of ${\mathcal{D}}_{2}$, however, does not change
significantly the structure of the wake within the Mach cone
for ${\mathcal{M}}_{\rm eff}\leq 1.5$. At larger Mach number
the contribution of ${\mathcal{D}}_{2}$ relative to 
${\mathcal{D}}_{1}$ becomes less and less 
important. Since the structure of 
${\mathcal{D}}_{1}$ is a scaled version of the wake
generated in the Newtonian case, the difference
between the structure of a wake generated by a small highly-supersonic
perturber in MOND, would be likely too subtle to be distinguished
from a perturber of fictitious mass $(1-T_{0})\mu_{0}^{-1}$ in Newtonian
gravity.
Differences only appear in the vicinity of the perturber.

\subsubsection{External field orthogonal to the velocity flow }
\label{sec:orthogonalsec}
Suppose now that the external field is along the $x$-axis. 
By denoting now 
$\hat{x}=x/r_{p}$, $\hat{y}=y/r_{p}$, $\hat{z}=z/r_{p}$, and
by reasoning entirely analogous to that leading to Eqs 
(\ref{eq:master1})-(\ref{eq:meff}), one finds
that when the external field is perpendicular to the
velocity flow, ${\mathcal{M}}_{\rm eff}$, $\hat{g}_{1}$ and $\hat{g}_{2}$ are
given by 
\begin{equation}
{\mathcal{M}}_{\rm eff}={\mathcal{M}},
\end{equation}
\begin{equation}
\hat{g}_{1}(\hat{\bmath{r}})=\left[1+\frac{\hat{x}^{2}}{1+L_{0}}+\hat{y}^{2}+\hat{z}^{2}\right]^{-5/2},
\end{equation}
and
\begin{equation}
\hat{g}_{2}(\hat{\bmath{r}})=
\left[\frac{2\hat{x}^{2}}{1+L_{0}}-\hat{y}^{2}-\hat{z}^{2}\right]\hat{g}_{1}.
\end{equation}
In this case, the dependence on $L_{0}$ does not factorize so that one
needs to calculate the integrals (\ref{eq:master2c}) for each pair 
$({\mathcal{M}}_{\rm eff},L_{0})$.

We found numerically that 
${\mathcal{D}}_{1}$ is very axisymmetric around the $x$ axis; 
the effect of the gravitational dilation on ${\mathcal{D}}_{1}$
is small.  For instance, when adopting
$L_{0}=1$, the angular variations of ${\mathcal{D}}_{1}$ are
less than $2\%$, $2\%$, $8\%$ and $13\%$ for ${\mathcal{M}}=0.4$,
${\mathcal{M}}=0.8$, 
${\mathcal{M}}=1.25$ and ${\mathcal{M}}=1.8$, respectively. 
${\mathcal{D}}_{1}$ is almost undistinguishable (differences of $\sim 5$ 
percent
or less) from its counterpart in the axisymmetric case, 
and thus they are not shown. In particular, the opening angle 
of the Mach cone for supersonic perturbers
in physical coordinates ($z,R$), 
is the same as in the axisymmetric case.

\begin{figure*}
\epsfig{file=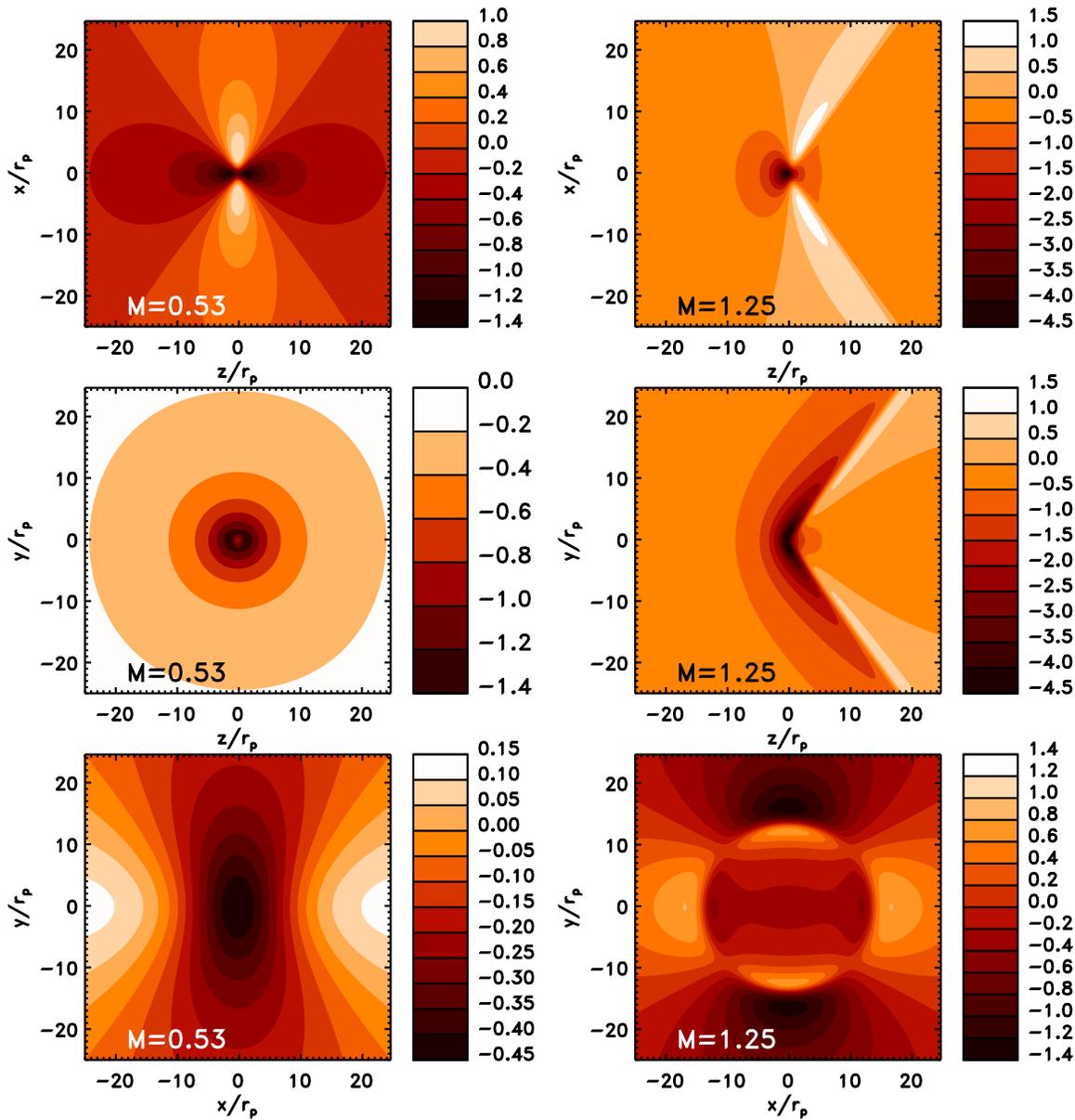,angle=0,width=17.5cm}
  \caption{Distribution of the dimensionless integral $\hat{I}_{2}$
at cut-off planes $y=0$ (upper), $x=0$ (middle) and $z=10r_{p}$ 
(lower) for ${\mathcal{M}}=0.53$ (left column)
and ${\mathcal{M}}=1.25$ (right column), in linear scale. 
The velocity of the perturber is along
the $z$-axis, whereas the external gravitational acceleration
is along the $x$-axis and $L_{0}=1$. }
\label{fig:nonaxi}
\end{figure*}

Unlike ${\mathcal{D}}_{1}$, ${\mathcal{D}}_{2}$ is expected to 
depart from axisymmetry about the $z$ axis.
We will focus again on the deep-MOND limit ($L_{0}=1$).
Figure \ref{fig:nonaxi} shows $\hat{I}_{2}$, which is
proportional to ${\mathcal{D}}_{2}$, in 
three perpendicular planes: $y=0$, $x=0$ and $z=10 r_{p}$, for
${\mathcal{M}}=0.53$ and ${\mathcal{M}}=1.25$.
When the perturber moves subsonically, the structure of ${\mathcal{D}}_{2}$ 
in the $y=0$ plane looks pretty much like in the axisymmetric case
after a rotation of $\pi/2$ (compare Fig.~\ref{fig:I2axi} 
and Fig.~\ref{fig:nonaxi}). In the $x=0$ plane, however, the density map
is notoriously different. It clearly shows that the configuration
$\hat{\mathcal{D}}_{2}$ has not an axial symmetry around the $z$ axis. 

For subsonic perturbers, ${\mathcal{D}}_{2}$ may be able to change
the flattening parameter of the wake.
As an example, consider ${\mathcal{M}}=0.8$ and $L_{0}=1$. In this situation,
${\mathcal{D}}_{1}$ has isodensity
contours with $q=0.6$. If the contribution of ${\mathcal{D}}_{2}$
is added, the composed wake exhibits $q=0.43$ in the $y=0$ plane, and
$q=0.54$ in the $x=0$ plane (not shown).

For a supersonic perturber, the overdensity regions in ${\mathcal{D}}_{2}$
are not located any longer at the head of the body; regions
at the front exhibit a decrease in density. Zones with
density depletion, that is ${\mathcal{D}}_{2}<0$, can be found 
at both downstream and upstream. 
The maximum density enhancement in ${\mathcal{D}}_{2}$
appears smaller, by a factor $\sim 4$ at ${\mathcal{M}}=1.25$, 
than in the axisymmetric configuration. 
Figure \ref{fig:nonaxi} shows the complexity of the structure in 
the $(x,y)$ plane (lower panels).  The isodensity contours in that 
plane turn up to be elongated along the $x$-axis.

\subsection{The gravitational drag on the body}
\label{sec:dragmain}
Once the structure of the wake ${\mathcal{D}}$ is constructed, 
it is straightforward
to evaluate the drag force exerted on the perturber by its wake. 
By symmetry, the only non-vanishing component lies along the $z$ axis.
In particular, the drag force in the axisymmetric case is
\begin{equation}
F_{DF}=2\pi \frac{GM\rho_{0}}{\mu_{0}}
\int\int dz dR R \frac{{\mathcal{D}}z}{((1+L_{0})R^{2}+z^{2}+\bar{r}_{p}^{2})^{3/2}}.
\end{equation}
If $F_{DF,1}$ and $F_{DF,2}$ denote the contribution to the drag force
by components ${\mathcal{D}}_{1}$ and ${\mathcal{D}}_{2}$, respectively,
we have:
\begin{equation}
F_{DF}=F_{DF,1}+F_{DF,2},
\end{equation}
\begin{equation}
F_{DF,1}=\frac{3}{2}\frac{1-T_{0}}{\mu_{0}^{2}(1+L_{0})}
\frac{G^{2}M^{2}\rho_{0}}{c_{\infty}^{2}}
{\mathcal{F}}_{1},
\label{eq:fdf1}
\end{equation}
\begin{equation}
F_{DF,2}=\frac{3}{2}\frac{T_{0}}{\mu_{0}^{2}(1+L_{0})}
\frac{G^{2}M^{2}\rho_{0}}{c_{\infty}^{2}}
{\mathcal{F}}_{2},
\end{equation}
where
\begin{equation}
{\mathcal{F}}_{i}=\int\int d\hat{z} d\hat{R}
\frac{\hat{I}_{i}\hat{R}\hat{z}}{(1+\hat{R}^{2}+\hat{z}^{2})^{3/2}}.
\label{eq:itf}
\end{equation}

We first estimate the relative contribution of $F_{DF,2}$ as compared
to $F_{DF,1}$. For supersonic perturbers,
${\mathcal{F}}_{2}$ and ${\mathcal{F}}_{1}$ were calculated 
by carrying out the integration in Eq.~(\ref{eq:itf})
over all our box domain.  Hence, the $\hat{z}$-integral 
has upper/lower limits $\hat{z}=\pm 25$ 
and the $\hat{R}$-integral has limits $\hat{R}=0,25$.
This implies that our steady-state wakes  
have an extent along $z$ of $z_{\rm max}\sim 25r_{p}$ to $35r_{p}$,
depending on $L_{0}$, and
a Coulomb logarithm $\ln\Lambda\sim \ln z_{\rm max}/2r_{p}\sim 3$
(S\'anchez-Salcedo \& Brandenburg 1999). 

For a purely steady-state density perturbation, the forward-backward 
symmetry in the subsonic case argues that zero net force acts
on the perturber.  However, Ostriker (1999) noticed that this
conclusion is misleading because, although complete ellipsoids exert no
net force on the perturber, there are always cut-off ones within
the sonic sphere that exert a gravitational drag.
Thus, we estimate the gravitational drag on
a subsonic perturber by integrating Eq.~(\ref{eq:itf}) 
over the largest sonic sphere
contained in our computational domain.

\begin{figure}
\epsfig{file=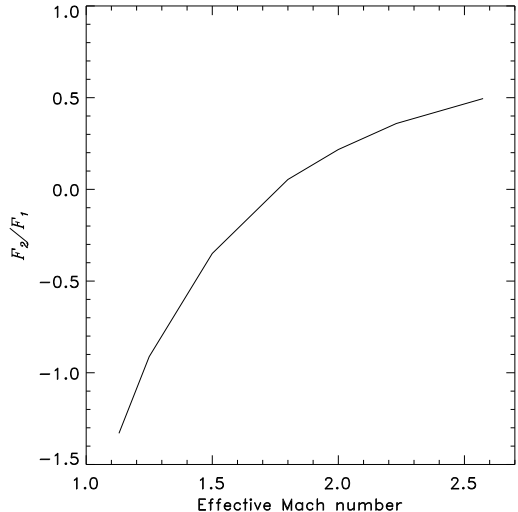,angle=0,width=7.5cm,height=7.4cm}
  \caption{Ratio between 
${\mathcal{F}}_{2}$ and ${\mathcal{F}}_{1}$, as defined in
Eq.~(\ref{eq:itf}), versus the
effective Mach number, for the axisymmetric case.
At ${\mathcal{M}}_{\rm eff}<1$,
the ratio depends on the adopted value for $L_{0}$ 
and hence is not shown.}
\label{fig:df}
\end{figure}

The ratio ${\mathcal{F}}_{2}/{\mathcal{F}}_{1}$ depends on
the angle between $\bmath{v}_{\infty}$ and $\bmath{g}_{\rm ext}$.
For illustration, Fig.~\ref{fig:df} shows its behaviour
as a function
of ${\mathcal{M}}_{\rm eff}$ for supersonic
bodies moving along the external gravitational field. 
In such a case, ${\mathcal{F}}_{2}$ is positive at large Mach numbers, 
implying that it contributes to drag
the body. It becomes zero at ${\mathcal{M}}_{\rm eff}\simeq 1.75$ and
negative for lower values. The relative contribution
of ${\mathcal{D}}_{2}$ becomes more important when approaching the transonic 
motion. 
At large Mach numbers, the ratio ${\mathcal{F}}_{2}/{\mathcal{F}}_{1}$
increases monotonically but very slowly.

\begin{figure}
\epsfig{file=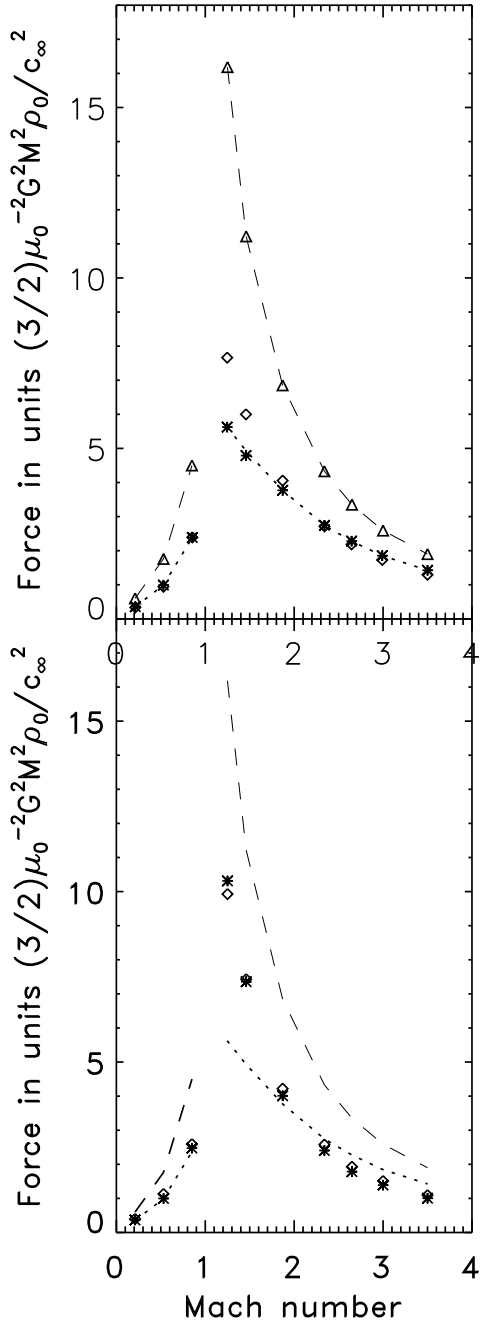,angle=0,width=13.5cm,height=20.4cm}
  \caption{DF force in a gaseous medium as a function of
the Mach number when $\bmath{v}_{\infty}$ and $\bmath{g}_{\rm ext}$
are parallel (top) and orthogonal (bottom).
The open diamonds correspond to the drag by the component
${\mathcal{D}}_{1}$ in the deep-MOND limit and the asterisks
correspond to the total drag. The dashed line and the triangles show the drag
in Newtonian gravity. To help comparison,
the total drag in the axisymmetric case and the Newtonian
drag have been also plotted in
the lower panel (dotted line and dashed line, respectively).
}
\label{fig:dfcomplete}
\end{figure}

In order to visualize the importance of including $F_{DF,2}$, 
Figure \ref{fig:dfcomplete} shows the total drag force exerted on the perturber 
as a function of ${\mathcal{M}}$, together with $F_{DF,1}$, 
in the deep-MOND limit ($L_{0}=1$, $T_{0}=1/6$). 
As anticipated, the sign of $F_{DF,2}$ depends on the angle
between $\bmath{v}_{\infty}$ and $\bmath{g}_{\rm ext}$. 
In contrast to the axisymmetric case,
when $\bmath{v}_{\infty}$ and $\bmath{g}_{\rm ext}$ are perpendicular,
$F_{DF,2}$ is positive for
transonic Mach numbers and becomes negative (reduces friction)
at high Mach numbers. 
The contribution to
the drag by the component ${\mathcal{D}}_{2}$ is more important
in the axisymmetric case but it is only noticeable ($>20\%$) 
at $1<{\mathcal{M}}<1.5$. 
Our numerical calculations show that 
$F_{DF,1}$ scales with the size of the box domain as 
$\propto \ln z_{\rm max}$, whereas
$F_{DF,2}$ increases somewhat slower with $z_{\rm max}$.
Therefore, the relative importance
of $F_{DF,2}$ is expected to be less  
for larger $z_{\rm max}$.

Now, we wish to compare the drag force in deep MOND ($L_{0}=1$) and in
Newtonian gravity ($L_{0}=0$). Figure \ref{fig:dfcomplete} also 
shows the Newtonian drag force experienced by a
body of mass $M$, travelling at Mach number ${\mathcal{M}}$ 
when the wake has the same extent
as in the MOND case, i.e.~$z_{\rm max}=25\sqrt{2} r_{p}$.
The drag force in MOND is a factor $\alpha \mu_{0}^{-2}$ 
larger than in Newton, where $\alpha$ is a form factor
that depends on the Mach number and on the angle between $\bmath{v}_{\infty}$
and $\bmath{g}_{\rm ext}$. In the axisymmetric
case, $\alpha\simeq 0.6$ at low Mach numbers 
(${\mathcal{M}}\lesssim 0.5$), and becomes
$\alpha\simeq 0.5$ at $0.5<{\mathcal{M}}\lesssim 1.0$.
For supersonic Mach numbers, 
$\alpha\simeq 0.4$ at $1<{\mathcal{M}}\lesssim 1.5$ and
increases monotonically with Mach number
up to $\simeq 0.8$ at high Mach numbers. 
The explanation for $\alpha=0.8$ at high Mach numbers is
covered in detail in the Appendix \ref{sec:alpha8}.
When the direction of motion is perpendicular to the external field,
$\alpha$ is very similar to its value in the axisymmetric case at
${\mathcal{M}}<1$. At supersonic velocities, $\alpha\simeq 0.6$
at $1<{\mathcal{M}}\lesssim 1.5$, and falls monotonically
down to $0.5$ at high Mach numbers.
Hence, the drag force may vary with the angle between
$\bmath{v}_{\infty}$ and $\bmath{g}_{\rm ext}$ by as much
as $50$ percent.

\section{Some implications}

\subsection{Wakes by galaxies and falling groups in clusters}
\begin{figure}
\epsfig{file=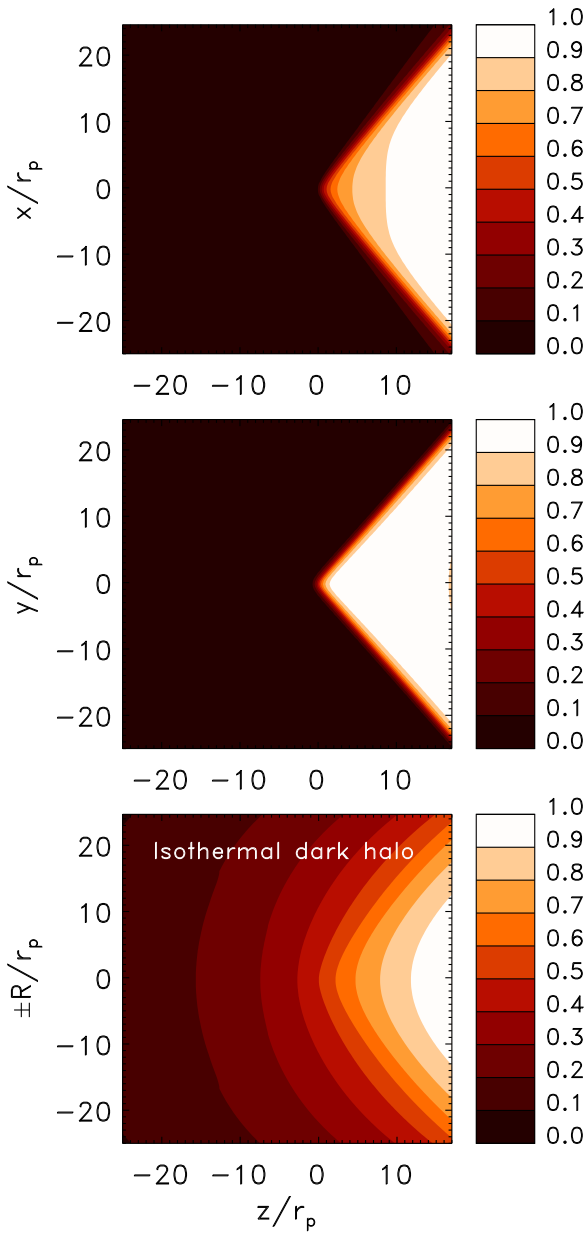,angle=0,width=17.5cm}
  \caption{Normalized X-ray surface brightness maps for a `compact'
body moving
at ${\mathcal{M}}=1.25$ in the deep-MOND non-axisymetric case for
a line-of-sight along the $y$-axis (upper panel) and along the
$x$-axis (middle panel), along with the X-ray map
for an `extended' pseudo-isothermal body in Newtonian gravity
(bottom).
The background field lays along
the $x$-axis. The observer's line of sight makes an angle
$\pi/2$ with respect the velocity of the perturber.}
\label{fig:xrays}
\end{figure}

In this section, we discuss the implications of X-ray observations of
the morphology of wakes in clusters of galaxies for 
modified gravities.  
As in Furlanetto \& Loeb (2002), let us assume
that the galaxy or group of galaxies  moves {\it supersonically}
through a constant density cluster core 
surrounded by a isothermal envelope:
\[ \rho_{IC}(r) = \left\{ \begin{array}{ll}
         \rho_{0} & \mbox{if $r \leq r_{c}$};\\
        2\rho_{0}/[1+(r/r_{c})^{2}] & \mbox{if $r \geq r_{c}$}.\end{array} \right. \] 
where $\rho_{0}$ is the density in the core and $r_{c}$ is the core
radius.
The emitted surface brightness $S=\int (\epsilon_{ff}/4\pi)dl$, where
$\epsilon_{ff}$ is the bremsstrahlung free-free volume emissivity, will
be enhanced in the wake by:
\begin{equation}
\frac{\delta S}{S}\sim 
\chi \frac{\delta \Sigma}{\Sigma},
\end{equation}
where $\Sigma$ is the column density, and $\chi \approx 4$ if the gas 
is isothermal, or $\approx 5$ if it varies adiabatically
(Furlanetto \& Loeb 2002).
Maps of X-ray for a ${\mathcal{M}}=1.25$ body with $\bmath{v}_{\infty}$
perpendicular to the external gravitational field are
shown in Fig.~\ref{fig:xrays}.
The structure of the emission in the wake 
is very similar to that formed by a ``compact'' 
gravitational perturber under Newton gravity. 
The projected X-ray emission
along line-of-sights perpendicular to the
direction of motion is roughly independent of the location in the wake, 
except near the edges of the cone, and is given by
\begin{equation}
\frac{\delta S}{S}\simeq \frac{2\pi\chi}{(2+\pi)} 
\frac{GM}{\mu_{0}\beta c_{\infty}^{2}r_{c}}.
\end{equation} 
As a consequence, in searching for the wake, one may expect an
abrupt jump in surface brightness at the edge of its cone.

We derived the gravitational wake induced by an extended body with
a Plummer profile having a fast density decay at large radii
($\rho \propto r^{-5}$) in an attempt to model the baryonic mass
of a certain bound object.  Wakes in the
cold dark matter (CDM) scenario are expected to be different than in
purely baryonic MOND because
the dark matter component in the halo of galaxies
decays as $r^{-2}$, much more slowly than the
baryonic mass density ($\leq r^{-3}$).
To illustrate this, Figure \ref{fig:xrays} also
shows the X-ray emission generated by a pseudo-isothermal perturber
with core radius $r_{p}$ in the Newtonian case.
We see that the X-ray
emission in the standard CDM scenario is more cuspy.
A MOND wake can be distinguished from a CDM wake by detecting 
a sharp X-ray enhancement along the Mach cone.
Furlanetto \& Loeb (2002) made a detailed
analysis of the wake morphology in CDM models for collisionless
and fully collisional (fluid) dark matter (FDM) in the supersonic case.
They found that, because in the collisional
case the dark halo is truncated by ram pressure stripping, the
X-ray emission of the wake is rather flat, similar to that we find
in MOND.
Due to the quantitative similarity between the wake in MOND and
in FDM, many of the observational suggestions raised by Furlanetto
\& Loeb to distinguish between FDM and CDM 
can be used to distinguish between dark matter or MOND.
We can repeat the reasoning of Furlanetto \& Loeb (2002) and argue
that the observations 
of the wake of the elliptical galaxy NGC 1404 in the core of the
Fornax group marginally support CDM
against MOND, but the evidence is very weak. In the last decade,
this type of observations has improved considerably (e.g.,
Drake et al.~2000; Neumann et al.~2001; Machacek et al.~2005, 2007;
Sun et al.~2006). Still,
it is difficult to draw some firm conclusions because of the difficulty
to isolate the structure of the gravitational wake
from the hydrodynamical wake, that is the mass in the wake stripped
from the own galaxy by ram pressure.
We must warn that MOND and FDM predict the same structure of the
gravitational wake past a galaxy but, in many other astrophysical aspects,
they must give different predictions since they are not equivalent.

In principle, observations of
the tails of subclusters of galaxies are a potential route
to distinguish betweeen collisionless CDM and MOND.
However, it is a well established issue that MOND still requires
dark matter at cluster scales (e.g., The \& White 1988; Sanders 1999). 
The inclusion of an isothermal dark matter component in MOND
erases somewhat the abovementioned differences between the MOND wake and
the CDM wake. The 
observed displacement between the X-ray peaks and the associated 
mass distribution, as derived from lensing
data in the Bullet Cluster, basically rules out FDM 
(Markevitch et al.~2004) but not necessarily MOND (Angus et al.~2007).

\subsection{DF timescale in a spherical system}
In \S \ref{sec:dragmain} we derived the DF force experienced by a body of
mass $M$ travelling on a rectilinear orbit through a homogeneous 
fluid medium in deep-MOND
and found that:
\begin{equation}
F_{DF}=-\frac{\alpha}{\mu_{0}^{2}}\frac{4\pi G^{2}M^{2}\rho_{0}}{v_{\infty}^{2}}
\ln\Lambda,
\label{eq:parametricforce}
\end{equation}
where $\alpha$ depends on the Mach number and on the angle that makes
the velocity of the perturber and the direction of the external
field. 
How does it compare with the DF force in a collisionless medium?
The formula for the collisionless MOND DF force was derived
by Ciotti \& Binney (2004). They show that it is similar to the Newtonian case
but replacing $G\rightarrow Ga_{0}/g_{\rm ext}$ plus an extra factor
of $\sqrt{2}$: 
\begin{equation}
F_{DF,{\rm no-col}}=-\frac{1}{\sqrt{2}\mu_{0}^{2}}
\frac{4\pi G^{2}M^{2}\rho_{0}}{v_{\infty}^{2}}
H \ln\Lambda,
\label{eq:ciotti04}
\end{equation}
where
\begin{equation}
H={\rm erf}\left(\frac{v_{\infty}}{\sqrt{2}\sigma}\right)
-\sqrt{\frac{2}{\pi}}\frac{v_{\infty}}{\sigma}
\exp\left(-\frac{v_{\infty}^{2}}{2\sigma^{2}}\right),
\end{equation}
with $\sigma$ the velocity dispersion of the Maxwellian distribution
of velocities of background particles
(e.g., Binney \& Tremaine 1987; S\'anchez-Salcedo et al.~2006).
By comparing Eqs (\ref{eq:parametricforce}) and (\ref{eq:ciotti04}),
we can generalize
the conclusion of Ostriker (1999) but for MOND: since the functional 
form of the gaseous DF
drag is much more sharply peaked near ${\mathcal{M}}=1$, the drag force
near ${\mathcal{M}}=1$ is larger in a gaseous medium than in a 
collisionless medium with the same density and $\sigma=c_{\infty}$. 
At ${\mathcal{M}}\approx 1$, using $\alpha\approx 0.5$ 
(see \S \ref{sec:dragmain})
there is factor of $\sim 3$ difference in the MOND force between
the two cases. For ${\mathcal{M}}<1$, the drag force is generally
larger in a collisionless medium than in a gaseous medium, because
pressure forces in a gaseous medium create a much more symmetric
density perturbation in the background.

For many problems of astrophysical interest,
it is convenient to have
the DF inspiraling timescale for a body that is initially on
a circular orbit.
For circular orbits, the above formula is correct
as long as the maximum impact parameter in the Coulomb logarithm 
is taken as $\sim 2R_{p}$, where $R_{p}$ is
the instantaneous orbital radius of the perturber 
(S\'anchez-Salcedo \& Brandenburg 2001; Kim \& Kim 2007).
Consider a massive body embedded in the gaseous outer
spherical envelope of a galaxy of mass $M_{G}$, with 
circular speed $v_{c}=(GM_{G}a_{0})^{1/4}$. 
Suppose that the gas is isothermal and in hydrostatic equilibrium
in the gravitational potential of the parent galaxy.
In a typical galaxy, the outer parts are expected to be
in the deep-MOND regime, hence $g_{\rm ext}=\sqrt{GM_{G}a_{0}}/r$
and $\mu_{0}=(GM_{G}/a_{0})^{1/2}r^{-1}$.
For the sake of clarity, let us assume that the gas has the
virial temperature: $c_{\infty}\simeq v_{c}/\sqrt{2}$. 
Imposing hydrostatic equilibrium, the unperturbed gas in the
envelope will pursue the following profile 
\begin{equation}
\rho_{g}=\rho_{s}\left(\frac{r}{r_{s}}\right)^{-2},
\label{eq:sphericalback}
\end{equation}
where $\rho_{s}$ is the gas density at the radius of reference $r_{s}$.
By substituting the values for $\mu_{0}$ and $\rho_{g}$ into 
Eq.~(\ref{eq:parametricforce}), and equating the rate of decrease of angular
momentum $d(M v_{c}r)/dt$ to the torque $rF_{DF}$, we find the
deep-MOND evolution of the orbital decay of a massive perturber's near-circular
orbit in a gaseous isothermal spherical distribution 
\begin{equation}
\frac{r}{r_{\rm init}}=\exp\left(-\frac{t-t_{\rm init}}{\tau_{M}}\right),
\end{equation}
where $r_{\rm init}$ is the orbital radius at $t=t_{\rm init}$ and
\begin{equation}
\tau_{M}=\frac{\mu_{s}^{2}}{\alpha}\frac{v_{c}^{3}}{4\pi G^{2}M}\frac{1}
{\rho_{s}\ln\Lambda},
\end{equation}
with $\mu_{s}=\mu_{0}(r_{s})$.

Consider now its equivalent Newtonian system (ENS) that is, the
Newtonian system (dark matter plus gas) in which the baryonic
gas has exactly the same density distribution as in the MOND system.
The density distribution of dark matter in the ENS is:
\begin{equation}
\rho_{dm}=\frac{v_{c}^{2}}{4\pi Gr^{2}}.
\end{equation}
In the ENS, the dark matter component added to have the same ``dynamics''
also contributes to the DF experienced by the body. The frictional
force in the ENS is 
\begin{equation}
F_{DF}=-\frac{4\pi G^{2}M^{2}(\rho_{g}+0.428\rho_{dm})}{v_{c}^{2}}
\ln\Lambda.
\end{equation}
The factor $0.428$ arises because the dark matter has been considered
collisionless (see, e.g., eq.~[7-25] of Binney \& Tremaine 1987). 
Equating again angular momentum loss with the torque, we find $\tau_{ENS}$,
the time that the body takes to reduce its distance a factor $e$ in the 
ENS:
\begin{equation}
\tau_{ENS}=\frac{0.43}{1+0.428{\mathcal{R}}_{g}}\left(\frac{r_{\rm init}^{2}}{r_{s}^{2}}\right)
\frac{v_{c}^{3}}{4\pi G^{2} M}\frac{1}{\rho_{s}\ln\Lambda},
\end{equation}
where ${\mathcal{R}}_{g}\equiv \rho_{dm}/\rho_{g}=$const.
The ratio between the DF timescale in MOND and in ENS is:
\begin{equation}
\frac{\tau_{M}}{\tau_{ENS}}
=4.6(1+0.428{\mathcal{R}}_{g})\mu_{\rm init}^{2}.
\end{equation}
Here $\mu_{\rm init}\equiv \mu_{0}(r_{\rm init})$ and we used 
that $\alpha\simeq 0.5$ for ${\mathcal{M}}=1.4$ (see \S
\ref{sec:dragmain}).
In terms of 
${\mathcal{R}}_{t,{\rm init}}$ defined as the ratio between 
the mass in dark matter
and the baryonic mass (stars plus gas) inside $r_{\rm init}$, we have
\begin{equation}
\frac{\tau_{M}}{\tau_{ENS}}
=4.6(1+0.428{\mathcal{R}}_{g})\frac{1}{[1+{\mathcal{R}}_{t,{\rm init}}]^{2}}.
\end{equation}
In dwarf protogalaxies, before the gas is turned into stars,
${\mathcal{R}}_{g}\simeq{\mathcal{R}}_{t}$, and for typical
values of dark matter contents in these systems (${\mathcal{R}}_{t}\simeq 20$),
the gaseous DF timescale in MOND is $10$ times shorter than in the ENS.
Condensed objects that form early (e.g., globular clusters) could spiral
into the centre of their host galaxy more rapidly than would be predicted in
the standard CDM haloes.

\section{Summary}
After a discussion about the Bondi-Hoyle problem, we have calculated the 
gravitational density wake of a perturber
moving through a uniform gaseous medium in MOND. 
A hydrodynamical treatment provides useful insight into the problem
of DF.  In order to
describe the response of the far-field medium to a small perturber, 
we focused on the case when the perturbation is dominated by a constant
external field.
The analytical knowledge of the wake structure in this simple
case is useful to interpret fully non-linear simulations
of the MOND dynamics of realistic systems.
The structure of the wake depends
on the angle between
the velocity vector and the external gravitational field.
We have considered two cases: the velocity of the perturber 
being either parallel or orthogonal to the external field. For an intermediate
case, the structure of the wake will lay somewhere between the
extremes described here. 
The density wake is decomposed into two linear contributions.
The morphology of the dominant contribution is a scaled version
of the Newtonian wake, being the MOND perturbed density a factor
$\mu_{0}^{-2}$ larger than in Newtonian gravity. The second contribution to 
the wake depends greatly
on the angle which the direction of motion makes with the external field.

The MOND DF force on the perturber induced by its own wake,
as a function of the Mach number
has been also derived and compared with the drag force 
in the Newtonian case.  
It is important to know the dependence of the drag force strength on Mach
number to study the circularization of orbits.
Our results confirm earlier analyses suggesting that the
DF force is higher in MOND than in Newtonian gravity. 
The DF drag is larger when the motion of the perturber is along
the external field, especially at Mach numbers between $1$--$1.5$. 
In the context of the sinking satellite problem, the recent claim
that the existence of an extended globular cluster population
in Fornax is problematic for MOND gains strength
(S\'anchez-Salcedo et al.~2006; Nipoti et al.~2008).
We find that MOND predicts an even faster orbital decay, especially 
when the satellite
arrives to the halo of a galaxy before the gas has turned into
stars.

In MOND we show that a supersonic perturber generates
a wake in the gas with a well-defined Mach cone in which the
the surface density increases substantially in a narrow region
and then flattens. Because CDM haloes extend to larger radii than
the baryonic mass, the wakes in the two models have significant
morphological differences. 

Inherent to the Chandrasekhar treatment of DF is the assumption
that the perturber interacts
with {\it unbound} particles. In MOND, this requirement immediately 
demands the inclusion of the external field. 
Once the background gravitational field is included, two-body
orbits of distant particles are unbound, the interaction 
of the perturber
with the medium is {\it quasi-Newtonian}, and the effects of distant
encounters can be simply added. 
This topic is closely related to studies of relaxation processes
in self-gravitating media. 
In a forthcoming paper we will
present a standard derivation of the DF drag in collisionless systems
in MOND.

\section*{acknowledgements}
The manuscript has benefitted from a prompt and thoughtful referee report.
I would like to thank Juan Maga\~na for useful discussions and
Alfredo D\'{\i}az and Alfredo Santill\'an
for their advice on some technical aspects. This work was partly supported
by CONACyT project CB2006-60526 and PAPIIT IN114107-3.

\appendix
\section{The Bondi-Hoyle radius in MOND}
\label{sec:appBondi}
For a point-like particle, 
the linear MOND field equation (Eq.~\ref{eq:m86}) is readily solved (Milgrom
1986): 
\begin{equation}
\Phi_{1}(\bmath{r})=-\frac{1}{\mu_{0} }
\frac{GM}{\sqrt{(1+L_{0})(x^{2}+y^{2})+z^{2}}},
\end{equation}
where we have supposed that the external field is along the $z$-axis.
We are now in a position to derive the Bondi-Hoyle radius $r_{BH,1}$ 
for the potential $\Phi_{1}$. We know, however, that $\Phi_{1}$
is not a good approximation in the vicinity of the particle. Nevertheless,
since $|\Phi_{1}|$ overestimates the depth of the potential
once we approach the perturber (i.e.~$|\Phi_{1}|\geq |\Phi_{\rm exact}|$,
where $\Phi_{\rm exact}$ is the exact potential for a point mass), 
the Bondi-Hoyle radius derived using the potential
$\Phi_{1}$ provides an upper limit to the exact 
Bondi-Hoyle radius.

In the rest frame of the perturber,
consider a streamline that begins at $x,y,z$ coordinates $(x,y,-\infty)$.
A straight streamline is susceptible to bending and shocking if the
velocity of the
gas on that streamline (adding the thermal velocity) is larger
than the maximum value of the escape velocity along that streamline:
\begin{equation}
\frac{2}{\mu_{0}(1+L_{0})^{1/2}}\frac{GM}{\sqrt{x^{2}+y^{2}}}>
v_{\infty}^{2}+c_{\infty}^{2},
\end{equation}
where we have assumed that $\bmath{v}_{\infty}$ and
$\bmath{g}_{\rm ext}$ 
lie both along the $z$-axis.
According to the above equation, all the streamlines that start 
with impact parameter $b$ such as:
\begin{equation}
b<\frac{2}{\mu_{0}(1+L_{0})^{1/2}}\frac{GM}{c_{\infty}^{2}
(1+{\mathcal{M}}^{2})}
\end{equation}
bend significantly and go through shocks. In analogy with the Newtonian
case, the MONDian Bondi-Hoyle radius is:
\begin{equation}
r_{BH,1}= \frac{1}{\mu_{0}(1+L_{0})^{1/2}}\frac{GM}{c_{\infty}^{2}
(1+{\mathcal{M}}^{2})}.
\end{equation}

If $\bmath{g}_{\rm ext}$ lies along the $x$-axis (so that $\bmath{g}_{\rm ext}$
and $\bmath{v}_{\infty}$ are perpendicular), the condition
for blending is:
\begin{equation}
\frac{2}{\mu_{0}}\frac{GM}{\sqrt{x^{2}+(1+L_{0})y^{2}}}>
v_{\infty}^{2}+c_{\infty}^{2}.
\end{equation}
The streamlines subject to terminate in a shock are
\begin{equation}
\left(x^{2}+(1+L_{0})y^{2}\right)^{1/2}<\frac{2}{\mu_{0}}
\frac{GM}{c_{\infty}^{2}
(1+{\mathcal{M}}^{2})}.
\end{equation}
We see that, due to the dilation along the axis of the background field,
the curvature of streamlines with the same impact parameter 
depends on their azimuthal angle in the $(x,y)$ plane.
Nevertheless, there is a guarantee that the response of the gas will 
be linear at any radius beyond $2r_{BH,1}$ where
\begin{equation}
r_{BH,1}= \frac{1}{\mu_{0}}\frac{GM}{c_{\infty}^{2}
(1+{\mathcal{M}}^{2})}.
\end{equation}

\section{The component ${\mathcal{D}}_{1}$ in the axisymmetric
case for a time-dependent perturbation}
\label{sec:appendixB}
We may be interested in the temporal evolution of the wake when the perturber
is placed at $t=0$.
The Newtonian case was studied by Ostriker (1999) who
found that the induced wake density for this finite-time perturbation
has the following analytic form 
\begin{equation}
{\mathcal{D}}_{N}(t)=\frac{\xi GM/c_{\infty}^{2}}{\sqrt{z^{2}
-\beta^{2}R^{2}}},
\end{equation}
where $\beta^{2}={\mathcal{M}}^{2}-1$ and
\[ \xi = \left\{ \begin{array}{ll}
         2 & \mbox{if ${\mathcal{M}}>1$, 
$R^{2}+(z-v_{\infty}t)^{2}>(c_{\infty}t)^{2}$},\\
           & \mbox{$z/R>\beta$, 
                                 and $z<\beta c_{\infty}t/{\mathcal{M}}$};\\
         1 & \mbox{if $R^{2}+(z-v_{\infty}t)^{2} < (c_{\infty}t)^{2}$};\\
         0 & \mbox{otherwise.} \end{array} \right. \] 
We can take advantage of this Newtonian result to 
get ${\mathcal{D}}_{1}(t)$ in MOND, at distances
larger than a few $r_{p}$, as:
\begin{eqnarray}
{\mathcal{D}}_{1}(t)&\simeq &\frac{1-T_{0}}{(1+L_{0})^{1/2}}
\frac{\mu_{0}^{-1}\xi G\hat{M}_{1}r_{p}^{2}/c_{\infty}^{2}}
{\sqrt{\hat{z}^{2}-\beta_{\rm eff}^{2}
\hat{R}^{2}}}\nonumber\\
&=& (1-T_{0})
\frac{\mu_{0}^{-1}
\xi GM/c_{\infty}^{2}}{\sqrt{z^{2}-
\beta^{2}R^{2}}}.
\label{eqn:isocon}
\end{eqnarray}
Here we used again that $\hat{M}_{1}= M/r_{p}^{3}$.

\section{The drag force in MOND and Newton at
high Mach numbers}
\label{sec:alpha8}
According to Eq.~(\ref{eq:fdf1}), the drag force experienced 
by a Plummer body in Newtonian gravity is:
\begin{equation}
F_{DF,N}=\frac{3}{2}
\frac{G^{2}M^{2}\rho_{0}}{c_{\infty}^{2}}
{\mathcal{F}}_{1}.
\label{eq:fdfn}
\end{equation}
Let us compare $F_{DF,N}$ 
with $F_{DF,1}$ at ${\mathcal{M}}\gg 1$ in the axisymmetric case. 
Note that at those Mach numbers $F_{DF,2}$ can be ignored.
Combining Eqs (\ref{eq:fdf1}) and (\ref{eq:fdfn}),
the drag force component $F_{DF,1}({\mathcal{M}},z_{\rm max})$ 
exerted on a body moving at ${\mathcal{M}}$ by a wake with
extent $z_{\rm max}$ can be recast:
\begin{equation}
F_{DF,1}
=\frac{1-T_{0}}{\mu_{0}^{2}(1+L_{0})}
F_{DF,N}\left({\mathcal{M}}_{\rm eff},\hat{z}_{\rm max}\right).
\label{eq:mmeff}
\end{equation}
In the limit ${\mathcal{M}}\gg 1$ and in the steady state, see e.g., Ostriker
(1999),
\begin{equation}
F_{DF,N}\propto \frac{1}{v_{\infty}^{2}}
\ln\left(\frac{z_{\rm max}}{r_{\rm min}}\right).
\end{equation}
Therefore, for $z_{\rm max}\gg r_{\rm min}$, we have $F_{DF,N}\propto
v_{\infty}^{-2}$ and 
\begin{eqnarray}
F_{DF,N}({\mathcal{M}}_{\rm eff})\simeq 
\nonumber
\left(\frac{{\mathcal{M}}}{{\mathcal{M}}_{\rm eff}}\right)^{2}
F_{DF,N}({\mathcal{M}})\\ 
\simeq (1+L_{0}) F_{DF,N}({\mathcal{M}}).
\end{eqnarray}
Substituting into Eq.~(\ref{eq:mmeff}), we find that the drag force exerted
by ${\mathcal{D}}_{1}$ in the supersonic axisymmetric MOND case, 
is about $(1-T_{0})\mu_{0}^{-2}=(5/6)\mu_{0}^{-2}$ larger 
than in Newtonian gravity.

\end{document}